\documentclass[aps,12pt,pra,showpacs,preprint,a4paper,]{revtex4}

\usepackage{amsmath}
\usepackage{graphicx}
\usepackage{subfig}
\usepackage{array}
\usepackage{amsfonts}
\usepackage{epsf}
\usepackage{epsfig}
\usepackage{amssymb}

\begin{document}

\title{Exact Solutions of the Klein-Gordon Equation for q-Deformed Manning-Rosen Potential via Asymptotic Iteration Method}
\author{\small Tapas Das}
\email[E-mail: ]{tapasd20@gmail.com}\affiliation{Kodalia Prasanna Banga High
 School (H.S), South 24 Parganas, 700146, India}

\begin{abstract}
Asymptotic iteration method (AIM) is used to find the exact analytical solutions of the one dimensional Klein-Gordon equation for the q-deformed Manning-Rosen potential with equal Lorentz vector and scalar potential. The bound state eigenfunctions are obtained in terms of the hypergeometric functions. Using the present results energy eigenvalues and corresponding eigenfunctions of the special cases like P\"{o}schl-Teller potential, Rosen-Morse potential and Eckart type potential are derived before concluding the work.
\end{abstract}


\maketitle

\newpage

\section{I\lowercase{ntroduction}}
It is well known that the Klein-Gordon equation[1-2] can always be reduced to a of Schr\"{o}dinger-type equation specially when the Lorentz scalar and vector potential are equal[3].Generally, Klein-Gordon equation is used to describe spin zero particles whereas the Dirac equation for the $\frac{1}{2}$ spin particles in relativistic quantum mechanics[4].The problem of finding the exact solutions of non-relativistic wave equation (Schr\"{o}dinger equation) as well as for relativistic wave equation (Klein-Gordon equation or Dirac equation), with different special potential models, have received focal attention in quantum mechanics[5-27]. Some authors [28-43] extended the investigations for arbitrary dimensions also. Such studies allow us to look deep inside the phenomenon that occurs in quantum chemistry, molecular, atomic as well as in hadronic physics.\\
Nowadays, Researchers have shown enormous interest in hyperbolic type potentials[5-9,24-25]. The short range Manning-Rosen potential[44] was proposed by M.F. Manning and N.Rosen in 1933 to explain the diatomic molecular model. This potential as well its modified forms have been used in several branches of physics for bound states and scattering properties of diatomic molecule. The use of its $q$-deformed version is quite useful in the sense that we can study different important other hyperbolic potentials just dialing appropriate $q$ with suitable coupling parameters of the potential.\\
In this research, the bound state energy eigenvalues and corresponding eigenfunctions of q-deformed Manning-Rosen potential with equal Lorentz scalar and vector potentials for the s-wave in the one dimensional Klein-Gordon equation have been studied via asymptotic iteration method [24-25,45-46]. The asymptotic iteration method is very powerful practical method which provides an accurate and efficient way to obtain the spectrum of many particles in relativistic and non relativistic quantum mechanics.\\
To make this paper self contained in the next section a brief outline of AIM is given. In section III bound state energy eigenvalues and the corresponding eigenfunctions have been derived for the q-deformed Manning-Rosen potential. Section IV is for the discussion, where some special cases such that P\"{o}schl-Teller potential, Rosen-Morse potential and Eckart type potential are discussed briefly. Finally, the section V serves the conclusion of the present work.       
\section{T\lowercase{he} A\lowercase{symptotic} I\lowercase{teration} M\lowercase {ethod} (AIM)}
AIM is proposed to solve the homogeneous linear second order differential equations of the form
\begin{eqnarray}
y_n^{''}(x)=\lambda_0(x)y_n^{'}(x)+s_0(x)y_n(x)\,,
\end{eqnarray}
where $\lambda_0(x)\neq 0$ and the prime denotes the derivative with respect to $x$. The variables, $\lambda_0(x)$ and $s_0(x)$ are sufficiently differentiable. To find a general solution to this equation, we differentiate Eq.(1) with respect to $x$ and find
\begin{eqnarray}
y_n^{'''}(x)=\lambda_1(x)y_n^{'}(x)+s_1(x)y_n(x)\,,
\end{eqnarray}
where
\begin{align}
\lambda_1(x)=\lambda_0^{'}(x)+s_0(x)+\lambda_{0}^2(x)\,, \nonumber\\
s_1(x)=s_0^{'}(x)+s_0(x)\lambda_0(x)\,.
\end{align}
Similarly, the second derivative of Eq.(1) provides
\begin{eqnarray}
y_n^{''''}(x)=\lambda_2(x)y_n^{'}(x)+s_2(x)y_n(x)\,,
\end{eqnarray}
where 
\begin{eqnarray}
\lambda_2(x)=\lambda_1^{'}(x)+s_1(x)+\lambda_0(x)\lambda_1(x)\,, \nonumber\\
s_2(x)=s_1^{'}(x)+s_0(x)\lambda_1(x)\,.
\end{eqnarray}
We can easily iterate Eq.(1) up to $(k+1)th$ and $(k+2)th$ derivatives, $k=1,2,3,....$\\
Therefore, we have
\begin{eqnarray}
y_n^{(k+1)}(x)=\lambda_{k-1}(x)y_n^{'}(x)+s_{k-1}(x)y_n(x)\,,\nonumber\\
y_n^{(k+2)}(x)=\lambda_k(x)y_n^{'}(x)+s_k(x)y_n(x)\,,
\end{eqnarray}
where 
\begin{eqnarray}
\lambda_k(x)=\lambda_{k-1}^{'}(x)+s_{k-1}(x)+\lambda_0(x)\lambda_{k-1}(x)\,,\nonumber\\
s_k(x)=s_{k-1}^{'}(x)+s_0(x)\lambda_{k-1}(x)\,,
\end{eqnarray}
which are called as the recurrence relation. 
Now from the ratio of the $(k+2)th$ and $(k+1)th$ derivatives, we have
\begin{eqnarray}
\frac{d}{dx}\ln[y_n^{(k+1)}(x)]=\frac{y_n^{(k+2)}(x)}{y_n^{(k+1)}(x)}=\frac{\lambda_k(x)[y_n^{'}(x)+\frac{s_k(x)}{\lambda_k(x)}y_n(x)]}{\lambda_{k-1}(x)[y_n^{'}(x)+\frac{s_{k-1}(x)}{\lambda_{k-1}(x)}y_n(x)]}\,.
\end{eqnarray}
For sufficiently large $k(>0)$, if
\begin{eqnarray}
\frac{s_k(x)}{\lambda_k(x)}=\frac{s_{k-1}(x)}{\lambda_{k-1}(x)}=\alpha(x)\,,
\end{eqnarray}
which is the ``asymptotic" aspect of the method, then Eq.(8)is reduced to
\begin{eqnarray}
\frac{d}{dx}\ln[y_n^{(k+1)}(x)]=\frac{\lambda_k(x)}{\lambda_{k-1}(x)}\,,
\end{eqnarray}
which yields
\begin{eqnarray}
y_n^{(k+1)}(x)=C_1\exp\left(\int{\frac{\lambda_k(x)}{\lambda_{k-1}(x)}} dx\right)=C_1\lambda_{k-1}(x)\exp\left(\int{[\alpha(x)+\lambda_0(x)]dx}\right)\,,
\end{eqnarray}
where $C_1$ is the integration constant and right hand side of the Eq.(11) is obtained by using Eq.(9) and Eq.(10). Inserting Eq.(11)
into Eq.(6), the first order differential equation is obtained as
\begin{eqnarray}
y_n^{'}(x)+\alpha(x)y_n(x)=C_1\exp\left(\int{[\alpha(x)+\lambda_0(x)]dx}\right)\,.
\end{eqnarray}
This is a first order differential equation which is very easy to solve and general solution of Eq.(1) can be obtained as:
\begin{eqnarray}
y_n(x)=\exp\left(-\int^x\alpha(x_1)dx_1\right)\left[C_2+C_1\int^x\exp\left(\int^{x_1}[\lambda_0(x_2)+2\alpha(x_2)]dx_2\right)dx_1\right]\,.
\end{eqnarray}
For a given potential, first the idea is to convert one dimensional Klein-Gordon equation to the form of Eq.(1) which gives $s_0(x)$ and $\lambda_0(x)$. Then, using the recurrence relations given by Eq.(7) parameters $s_k(x)$ and $\lambda_k(x)$ are obtained. The termination condition of the method in Eq.(9) can be arranged as
\begin{eqnarray}
\Delta_k(x)=\lambda_k(x)s_{k-1}(x)-\lambda_{k-1}(x)s_k(x)=0\,,
\end{eqnarray}
where $k$ is the iteration number. For the exactly solvable potentials, the energy eigenvalues are obtained from the roots of Eq.(14) and the radial quantum number $n$ is equal to the iteration number $k$ for this case.For nontrivial potentials that have no exact solutions, for a specific $n$ principle quantum number, we choose a suitable $x_0$ point, determined generally as the maximum value of the asymptotic wave function or the minimum value of the potential [45] and the approximate energy eigenvalues are obtained from the roots of Eq.(14) for sufficiently great values of $k$ with iteration for which $k$ is always greater than $n$ in these numerical solutions.\\
The general solution of Eq.(1) is given by Eq.(13). The first part of Eq.(13) gives the polynomial solutions that are convergent and physical, whereas the second part of Eq.(13) gives non-physical solutions that are divergent. Although Eq.(13) is the general solutions of Eq.(1), we take the coefficient of the second part $C_1=0$, in order to find the square integrable solutions. Therefore, the corresponding eigenfunctions can be derived from the following wave function generator for exactly solvable potentials:
\begin{eqnarray}
y_n(x)=C_2\exp\left(-\int^x\frac{s_n(x_1)}{\lambda_n(x_1)}dx_1\right)\,,
\end{eqnarray}
where $n$ represents the principle quantum number.

\section{B\lowercase{ound} s\lowercase{tate} s\lowercase{olutions}}
In natural unit ($\hbar=c=1$) the one dimensional time independent K-G equation for a spinless particle of rest mass $m$ is written as
\begin{eqnarray}
\frac{d^2}{dx^2}\Psi(x)+\left[(E_n-V(x))^2-(m+S(x))^2\right]\Psi(x)=0\,,
\end{eqnarray}  
where $E_n$ represents the $n$ th state relativistic energy of the particle. $V(x)$ and $S(x)$ are the Lorentz vector and scalar potential respectively. Assuming $V(x)=S(x)$ we have
\begin{eqnarray}
\frac{d^2}{dx^2}\Psi(x)+\left[(E_{n}^2-m^2)-2( E_n+m)V(x)\right]\Psi(x)=0\,.
\end{eqnarray}
Now the $q$-deformed Manning-Rosen potential is of the form
\begin{eqnarray}
V_q(x)=V_1cosech_{q}^2(\alpha x)+V_2coth_q(\alpha x)\,,\;\,-1\leq q<0  \;\mbox{or}\; q>0\,,
\end{eqnarray}
where screening parameter $\alpha$ determines the range of the potential and $V_1 , V_2 $ are the coupling parameters describe the depth of the potential well. In general $q$-deformed hyperbolic functions are defined as
\begin{eqnarray*}
sinh_q(y)=\frac{1}{cosech_q(y)}=\frac{e^y-qe^{-y}}{2}\,\,, cosh_q(y)=\frac{e^y+qe^{-y}}{2}\,\,, coth_q(y)=\frac{cosh_q(y)}{sinh_q(y)}\,.
\end{eqnarray*}  
Using the above definition of $q$-deformed hyperbolic function, the potential function given by Eq.(18) can be rewritten as
\begin{eqnarray}
V(x)=4V_1\frac{e^{-2\alpha x}}{(1-qe^{-2\alpha x})^2}+V_2\frac{1+qe^{-2\alpha x}}{1-qe^{-2\alpha x}}\,.
\end{eqnarray}
Introducing the new variable $s=e^{-2\alpha x}$ and using Eq.(19), it is easy to write the Eq.(17) as
\begin{eqnarray}
\frac{d^2\Psi(s)}{ds^2}+\frac{1}{s}\frac{d\Psi(s)}{ds}+\left[-\frac{\epsilon_{n}^2}{s^2}-\frac{\gamma(\gamma-1)q}{s(1-qs)^2}-\frac{\beta^2(1+qs)}{s^2(1-qs)}\right]\Psi(s)=0\,,
\end{eqnarray}
where 
\begin{eqnarray}
\epsilon_{n}^2=\frac{m^2-E_{n}^2}{4\alpha^2}\,\,, \gamma(\gamma-1)q=\frac{2(E_n+m)V_1}{\alpha^2}\,\,,\beta^2=\frac{2(E_n+m)V_2}{2\alpha^2}\,.
\end{eqnarray}
To solve Eq.(20) by using AIM, we assume the following physical wave function satisfying the boundary conditions $\Psi(s=0)=0$ and $\Psi(s=\frac{1}{q})=0$
\begin{eqnarray}
\Psi(s)=s^c(1-qs)^{\gamma}f_n(s)\,,
\end{eqnarray}  
where $c=\sqrt{\epsilon_{n}^2+\beta^2}$.\\
Inserting the above wave function, Eq.(20) gives the following linear second order homogeneous differential equation 
\begin{eqnarray}
\frac{d^2}{ds^2}f_n(s)=\left\{\frac{qs(2c+2\gamma+1)-(2c+1)}{s(1-qs)}\right\}\frac{d}{ds}f_n(s)+\left\{\frac{2q\beta^2+2cq\gamma+q\gamma^2}{s(1-qs)}\right\}f_n(s)\,.
\end{eqnarray}
Now it is easy to find the solution of Eq.(23) using AIM. Comparing with Eq.(1) we have 
\begin{align}
\lambda_0(s)=\frac{qs(2c+2\gamma+1)-(2c+1)}{s(1-qs)}=\frac{qsk-u}{s(1-sq)} \nonumber\\
s_0(s)=\frac{2q\beta^2+2cq\gamma+q\gamma^2}{s(1-qs)}=\frac{ql}{s(1-sq)}\,,
\end{align}
where 
\begin{align}
k&=2c+2\gamma+1\,, &    u&=2c+1\,, &  l&=2\beta^2+2c\gamma+\gamma^2\,.
\end{align}
We may calculate $\lambda_k(s)$ and $s_k(s)$ from the recursion relation given by Eq.(7). This gives
\begin{eqnarray}
\lambda_1(s)=\lambda_{0}^{'}+s_0+\lambda_{0}^2=\frac{s^2q^2(k^2+k-l)+qs(l-2u-2ku)+u^2+u}{s^2(1-qs)^2}\nonumber\\
s_1=s_{0}^{'}+s_0\lambda_0=\frac{sq^2(2l+2k)-q(l+lu)}{s^2(1-qs)^2}........etc.
\end{eqnarray}
Here prime denotes the derivative with respect to $s$. Now Eq.(14) gives
\begin{eqnarray}
\Delta_1(s)=s_0(s)\lambda_1(s)-s_1(s)\lambda_0(s)=q^2\frac{l(l+k)}{s^2(1-qs)^2}\,.
\end{eqnarray}
The root of Eq.(27) gives the first value of $\epsilon_{n}^2$ i.e $\epsilon_{0}^2$. Similarly finding other $\Delta_n(s)$ we can explore different $\epsilon_{n}^2$. That means
\begin{eqnarray}
\Delta_1(s)=s_0(s)\lambda_1(s)-s_1(s)\lambda_0(s)=0\Rightarrow\epsilon_{0}^2=\frac{\gamma^4+4\beta^4}{4\gamma^2}\,, \nonumber\\
\Delta_2(s)=s_1(s)\lambda_2(s)-s_2(s)\lambda_1(s)=0\Rightarrow\epsilon_{1}^2=\frac{(\gamma+1)^4+4\beta^4}{4(\gamma+1)^2}\,,\nonumber\\
\Delta_3(s)=s_2(s)\lambda_3(s)-s_3(s)\lambda_2(s)=0\Rightarrow\epsilon_{2}^2=\frac{(\gamma+2)^4+4\beta^4}{4(\gamma+2)^2}\,,
\end{eqnarray}
.......and so on. Now using mathematical induction we can write the eigenvalues of the form
\begin{eqnarray}
\epsilon_{n}^2=\frac{(\gamma+n)^4+4\beta^4}{4(\gamma+n)^2}\,\,; n=0,1,2,3......
\end{eqnarray}
The energy eigenvalues of the $q$-deformed Manning-Rosen potential can be found from Eq.(21) as
\begin{eqnarray}
E_{n}^2=m^2-\alpha^2\left\{(\gamma+n)^2+\frac{4\beta^4}{(\gamma+n)^2}\right\}\,,
\end{eqnarray}
where $\gamma=\frac{1}{2}\pm\frac{1}{2}\sqrt{1+\frac{8(E_n+m)V_1}{q\alpha^2}}$.\\
It is to be noted that, by taking $qs=z$ the Eq.(23) transforms to the differential equation
\begin{eqnarray}
z(1-z)\frac{d^2G}{dz^2}+[u-z(v+w+1)]\frac{dG}{dz}-vwG=0\,,
\end{eqnarray}
where $v+w=2(c+\gamma)$ , $vw=2\beta^2+2c\gamma+\gamma^2=l$ , $u=2c+1$ and $G(z)\Rightarrow f_n(s)$.\\
Eq.(31) is satisfied by the hypergeometric function [47] $\,_{2}F_{1}(u,v,w;z)$.\\
Now we can derive the unnormalized eigenfunctions by using the wave function generator given by Eq.(15)
\begin{eqnarray}
f_n(s)=(-1)^nC_2\frac{\Gamma(n+2c+1)}{\Gamma(2c+1)}\,_{2}F_{1}(-n,2(c+\gamma)+n,1+2c;qs)\,,
\end{eqnarray}
where $\Gamma$ and $\,_{2}F_{1}$ are known as the gamma and the hypergeometric functions respectively[47].
Finally using Eq.(32) and Eq.(22) we can write the total unnormalized wave function as
\begin{eqnarray}
\Psi(s)=Ns^c(1-qs)^{\gamma}\,_{2}F_{1}(-n,2(c+\gamma)+n,1+2c;qs)\,, 
\end{eqnarray}
where $N$ is the normalization constant.
\section{D\lowercase{erivation} \lowercase{of} s\lowercase{pecial} c\lowercase{ases}}  
\begin{enumerate}
\item{\bf{ P\"{o}schl-Teller potential}}\\
Inserting $q=-1$, $V_1\Rightarrow -V_1$ and $V_2=0$ the potential given by Eq.(18) [or Eq.(19)] becomes 
\begin{eqnarray}
V(x)=-4V_1\frac{e^{-2\alpha x}}{(1+e^{-2\alpha x})^2}=-V_1sech^2\alpha x\,.
\end{eqnarray}
Immediately this helps to find the energy eigenvalues from Eq.(30) as
\begin{eqnarray}
E_{n}^2=m^2-\alpha^2(\gamma+n)^2\,,
\end{eqnarray}
where $\gamma=\frac{1}{2}\pm\frac{1}{2}\sqrt{1+\frac{8(E_n+m)V_1}{\alpha^2}}$. Now as $\beta\Rightarrow 0$ we have the constant $c=\epsilon_n$. This provides the eigenfunctions from Eq.(33) 
\begin{eqnarray}
\Psi(s)=Ns^{\epsilon_n}(1+s)^{\gamma}\,_{2}F_{1}(-n,2(\epsilon_n+\gamma)+n,1+2\epsilon_n,-s)\,.
\end{eqnarray}
These results are consistent with the work listed in reference [20].
\item{\bf{ Rosen-Morse potential}}\\
Inserting $q=-1$ and $V_1\Rightarrow -V_1$ the potential given by Eq.(18) [or Eq.(19)] becomes
\begin{eqnarray}
V(x)=-V_1sech^2\alpha x+V_2tanh \alpha x=-4V_1\frac{e^{-2\alpha x}}{(1+e^{-2\alpha x})^2}+V_2\frac{1-e^{-2\alpha x}}{1+e^{-2\alpha x}}\,.
\end{eqnarray}
As previous the energy eigenvalues for this case come out from Eq.(30) as 
\begin{eqnarray}
E_{n}^2=m^2-\alpha^2\left\{(\gamma+n)^2+\frac{4\beta^4}{(\gamma+n)^2}\right\}\,,
\end{eqnarray}
where $\gamma=\frac{1}{2}\pm\frac{1}{2}\sqrt{1+\frac{8(E_n+m)V_1}{\alpha^2}}$.\\
Corresponding eigenfunctions are
\begin{eqnarray}
\Psi(s)=Ns^c(1+s)^{\gamma}\,_{2}F_{1}(-n,2(c+\gamma)+n,1+2c,-s)\,, 
\end{eqnarray} 
were $c=\sqrt{\epsilon_n^{2}+\beta^2}$. \\
These results are similar to the work that listed in reference [25,51].
\item{\bf{ Eckart type potential}}\\
Inserting $q=1$ and $V_2\Rightarrow -V_2$ the potential given by Eq.(18) [or Eq.(19)] becomes
\begin{eqnarray}
V(x)=V_1cosech^2\alpha x-V_2coth\alpha x=4V_1\frac{e^{-2\alpha x}}{(1-e^{-2\alpha x})^2}-V_2\frac{1+e^{-2\alpha x}}{1-e^{-2\alpha x}}\,.
\end{eqnarray}
Again Eq.(30) gives the energy eigenvalues 
\begin{eqnarray}
E_{n}^2=m^2-\alpha^2\left\{(\gamma+n)^2+\frac{4\beta^4}{(\gamma+n)^2}\right\}\,,
\end{eqnarray}
where $\gamma=\frac{1}{2}\pm\frac{1}{2}\sqrt{1+\frac{8(E_n+m)V_1}{\alpha^2}}$.\\
Corresponding eigenfunctions are
\begin{eqnarray}
\Psi(s)=Ns^c(1-s)^{\gamma}\,_{2}F_{1}(-n,2(c+\gamma)+n,1+2c,s)\,, 
\end{eqnarray} 
were $c=\sqrt{\epsilon_n^{2}+\beta^2}$. 
\end{enumerate}
These results are also consistent with the work listed in reference [51].
\section{C\lowercase{onclusions}}
In this article, one dimensional Klein-Gordon equation has been solved for $q$-deformed Manning-Rosen potential with equal Lorentz Vector and scalar potential via AIM. The eigenfunctions are presented by hypergeometric functions. Some special cases such as P\"{o}schl-potential, Rosen-Morse potential, Eckrat type potential are also discussed in this work. We found that the results are in good agreement with the other findings in the literature.\\
It is evident that the AIM is a powerful, efficient and accurate alternative method of deriving energy eigenvalues and eigenfunctions of the hyperbolic type potentials that are analytically solvable. It should be mentioned that the AIM provides a closed form solutions for the exactly solvable problems. However, if there is no such a solutions, the results are obtained by using iterative approach[48-50].The results are sufficiently accurate for such special potentials at least for practical purpose.


\newpage
\begin{thebibliography}{99}
\bibitem{ref1} O.Z.Klein, Phys.{\bf37}, 895 (1926).

\bibitem{ref2} W.Z.Gordon, Phys.{\bf40},117 (1927).

\bibitem{ref3} R.L.Hall, Phys.Lett.A.{\bf372},12 (2007).

\bibitem{ref4} L.I.Schiff,\textit{Quantum Mechanics}, 3rd edition (McGraw-Hill, New York:1955).

\bibitem{ref5} W.C.Qiang, K.Li,W.L.Chen, J. Phys. A. Math. Theor.{\bf42}, 205306 (2009).

\bibitem{ref6} S.H.Dong, Commun.Theor.Phys. {\bf55}, 969 (2011).

\bibitem{ref7} K.J.Oyewumi, C.O.Akoshile, Eur. Phys. J. A {\bf45}, 311 (2010).

\bibitem{ref8} J.P.kllingbeck, A.Grosjean, G.Jolicard, J.Chem. Phys.{\bf116},447 (2002).

\bibitem{ref9} C.Berkdemir, A.Berkdemir, R.Sever, J.Phys.A: Math.Gen.{\bf39},13455 (2006).

\bibitem{ref10} W.Lucha, F.F.Sch\"{o}berl, Int.J.Mod.Phys.C {\bf10},607 (1999).

\bibitem{ref11} B.Roy, P.Roy, J.Phys.A: Math.Gen.{\bf35},3961 (2002).

\bibitem{ref12} L.Serra,E.Lipparini, Europhys.Lett.{\bf40}, 667 (1997).

\bibitem{ref13} V.Milanovic, Z.Ikovic, J.Phys.A.{\bf32}, 7001 (1999).

\bibitem{ref14} A.D.Alhaidari, Phys.Rev.A.{\bf66},042116 (2002).

\bibitem{ref15} F.Dominguez-Adame.Phys.Lett.A.{\bf136}, 175 (1989).

\bibitem{ref16} Z.Min-Chang, W.Zhen-Bang, Chinese Phys.Lett {\bf22}, 2994 (2005).

\bibitem{ref17} Z.Qiang, Y.Ping, G.Lun-Xun, Chinese Phys.{\bf15}, 35 (2006).

\bibitem{ref18} F.Yasuk, A.Durmus, I.Boztosun, J.Math.Phys.{\bf47}, 082302 (2006). 

\bibitem{ref19} Z.Min-Chang, W.Zhen-Bang, Chinese Phys.Lett {\bf16}, 1863 (2007). 

\bibitem{ref20} K.J.Oyewumi, T.T.Ibrahim, S.O.Ajibola, D.A.Ajadi, J.Vec.Rel.{\bf5}, 19 (2010).

\bibitem{ref21} W.C.Qiang, Chin.Phys.{\bf12}, 136 (2003).

\bibitem{ref22} W.C.Qiang, Chin.Phys.{\bf13}, 575 (2004).

\bibitem{ref23} J.Y.Guo, J.Meng, F.X.Xu, Chin.Phys.Lett, {\bf20(5)}, 602 (2003).

\bibitem{ref24} F.Taskin, I.Boztosun, O.Bayrak, Int.J.Theor.Phys.{\bf47}, 1612 (2008).

\bibitem{ref25} S.Debnath, B.Biswas, EJPT {\bf9(26)}, 191 (2012).

\bibitem{ref26} I.O.Vakarchuk,J.Phys.A.Math.Gen.{\bf38}, 4727 (2005).

\bibitem{ref27} A.N.Ikot, E.Maghsoodi, S.Zarrinkamar, H.Hassanabadi,{\bf18(1)}, 1003 (2014).

\bibitem{ref28} H.Hassanabadi, S.Zarrinkamar, A.A.Rajabi, Commun.Theor.Phys.{\bf55},541 (2011).

\bibitem{ref29} D.Agboola, Chin.Phys.Lett. {\bf27},040301 (2010).

\bibitem{ref30} S.H.Dong, G.H.Sun, Phys. Lett.A.{\bf314}, 261 (2003).
 
\bibitem{ref31} S.H.Dong, Phys.Scr.{\bf65}, 289 (2002).

\bibitem{ref32} S.M.Ikhdair, R.Server, Int.J.Mod.Phys.C {\bf18}, 1571 (2007).

\bibitem{ref33} D.Agboola,Phys.Scr.{\bf80}, 065304 (2009).

\bibitem{ref34} L.Y.Wang, X.Y.Gu,Z.Q.Ma,S.H.Dong,Found.Phys.Lett.{\bf15}, 569 (2002).

\bibitem{ref35} D.Agboola, Phys.Scr.{\bf81}, 067001 (2010).

\bibitem{ref36} D.Agboola, ACTA PHYSICA POLONICA A.{\bf120}, 371 (2011).

\bibitem{ref37} K.J.Oyewumi,F.O.Akinpelu,A.D.Agboola,Int.J.Theor.Phys.{\bf47}, 1039 (2008).  

\bibitem{ref38} G.R.Khan, Eur.Phys.J.D {\bf53}, 123 (2009).
 
\bibitem{ref39} S.Ortakaya, Chin.Phys.B.{\bf22(7)}, 070303 (2013).

\bibitem{ref40} M.Hamzavi, S.M.Ikhdair, K.E.Thylwe, Chin.Phys.B {\bf22(4)}, 040301 (2013).

\bibitem{ref41} T.Das, arXiv:1406.6282v1 [quant-ph],(2014).  

\bibitem{ref42} T.Das, A.Arda, arXiv:1308.5295v2 [math-ph],(2014).

\bibitem{ref43} T.Das, arXiv:1408.6139v1 [quant-ph],(2014). 

\bibitem{ref44} M.F.Manning, N.Rosen, Phys.Rev.{\bf44}, 953 (1933).

\bibitem{ref45} H.Ciftci, R.L.Hall, N.Saad, Phys.A:Math .Gen.{\bf36}, 11807 (2003).

\bibitem{ref46} H.Ciftci, R.L.Hall, N.Saad, Phys.A:Math .Gen.{\bf38}, 1147 (2005).

\bibitem{ref47} G. B. Arfken and H. J. Weber, \textit{Mathematical Methods for Physicists}, (Academic Press,San Diego: 1995).

\bibitem{ref48} F.M.Fernandez, J.Phys.A: Math.Gen. {\bf37}, 6173 (2004).

\bibitem{ref49} T.Barakat,Phys.Lett.A. {\bf344}, 411 (2005).

\bibitem{ref50} T.Barakat,J.Phys.A:Math.Gen.{\bf36}, 823 (2006).

\bibitem{ref51} S.M.Ikhdair, Journal of Quantum Information Science.{\bf1},73 (2011).
\end{thebibliography}
\end{document}